# Properties of pulsed-laser deposited nanocomposite NiO:Au thin films for gas sensing applications

I. Fasaki, M. Kandyla, M. Kompitsas[*]

National Hellenic Research Foundation, Theoretical and Physical Chemistry Institute, Vasileos Konstantinou Ave. 48, 11635 Athens, Greece

**Abstract**

Nanocomposite thin films formed by gold nanoparticles embedded in a nickel oxide matrix have been synthesized by a new variation of the pulsed laser deposition technique. Two actively synchronized laser sources, a KrF excimer laser at 248 nm and an Nd:YAG laser at 355 nm, were used for the simultaneous ablation of nickel and gold targets in oxygen ambient. The structural, morphological, and electrical properties of the obtained nanocomposite films were investigated in relation to the fluence of the laser irradiating the gold target. The nanocomposite thin films were tested as electrochemical hydrogen sensors. It was found that the addition of the gold nanoparticles increased the sensor sensitivity significantly.

[*] Corresponding author:
M. Kompitsas
National Hellenic Research Foundation
48, Vas. Konstantinou Ave
11635 Athens, Greece
Tel.  +302107273834
Fax  +302107273794,
E-mail: mcomp@eie.gr
URL: www.laser-applications.eu

# 1. Introduction

Nanocomposite metallic oxide thin films with Au nanoparticles find extensive applications, including among others: electrochemical $SnO_2$:Au sensors for CO with increased sensitivity [1]; biochemical sensors with enhanced response [2]; optical detection of CO based on absorbance changes of NiO:Au compounds [3]; gas detection selectivity between CO and $H_2$, by means of optical sensing based on CuO:Au compounds [4] and ozone detection by conductometric $SnO_2$:Au based gas sensors [5].

NiO, one of the very few known p-type metal-oxide semiconductors [6], is a dielectric, antiferromagnetic, electrochromic [7, 8], and catalytic [9] material with excellent chemical stability [10] and good gas sensing properties [11, 12]. NiO thin films have been prepared in the past by several methods, such as sputtering [12], chemical spray pyrolysis [13], and Pulsed Laser Deposition (PLD) [7, 8, 14, 15].

The nanocomposite system NiO:Au has been developed previously by pyrolysis [4, 16]. The nanoparticles had a diameter of a few tens of nm and caused a catalytic enhancement on the optical detection of CO. Recently, a new thin-film chemical growth process has been proposed and realized [17] that led to a device operating as a solid-state gas sensor with enhanced performance. Metallic (Au, Ag, Pt) nanoparticles served as nanoelectrodes between $SnO_2$ crystallites to reduce the resistivity of the device and facilitate gas detection using electrochemical sensing.

In Ref. [17], the metallic nanoparticles have been deposited either below ("bottom layout") or on the surface ("top layout") of the functional $SnO_2$ film. In this paper, we report on the growth of NiO:Au nanocomposite thin films, where Au nanoparticles are stochastically embedded inside the NiO matrix ("intermediate layout"). The setup we use has been developed previously [18] as an extension of the typical PLD technique. It consists of a synchronized dual-laser, dual-target system that allows the control of the concentration of any element into any metallic oxide matrix. This is achieved by adjusting the fluence of the laser incident on the *impurity* (additional element) target during film growth. In this way, the structural and electrical properties of the synthesized nanocomposite materials may be efficiently optimized for specific applications.

In the present study, the grown nanocomposites have been employed as hydrogen sensors. We manage to detect hydrogen efficiently for operating temperatures much lower than those found in the literature [12, 13, 16], in order to reduce the energy consumption of the sensor without sacrificing sensor sensitivity. Hydrogen sensing is important for many applications because hydrogen becomes explosive in air with a lowest explosion limit (LEL) of 4% (40,000 ppm).

2. **Materials preparation and methods**

The versatile Reactive Pulsed Laser Deposition (R-PLD) technique consists of a 248 nm KrF excimer laser for the ablation of the metallic Ni target and a 355 nm Q-switched Nd:YAG laser for the ablation of the Au foil. The two laser systems are actively synchronized and driven at 10 Hz by a common pulse generator. The KrF ("laser 1") laser fluence was fixed at 5.5 J/cm$^2$. The Nd:YAG ("laser 2") laser fluence was varied and set at 0, 1.3, 2.4, and 4.3 J/cm$^2$, in order to control the amount of Au embedded into the NiO martix. This way, four different sets of samples were prepared with varying Au concentrations. To avoid fast drilling, both targets were placed on a vacuum-compatible, computer controlled *XY* translation stage (Figure 1). Thermally oxidized Si substrates were positioned 40 mm away from the targets, in order for the two ablation plasmas to intersect on the substrate surface, as shown in Figure 1. The substrates were heated at 400 °C during thin film growth. The target irradiation time for every film was 90 min. Prior to each irradiation, the vacuum chamber was evacuated down to a residual pressure of $10^{-4}$ Pa. All depositions occurred in a dynamic 40 Pa oxygen pressure in order to obtain NiO.

The morphological and structural characteristics of the obtained films were investigated by Scanning Electron Microscopy (SEM-EDS) and by an X-Ray Diffraction (XRD) Siemens D 5000 instrument in the 35º-50º range. The resistivity of the films was measured using the four-point Van der Pauw method.

Hydrogen sensing has been investigated inside an aluminium vacuum chamber [18]. The chamber was initially evacuated down to 1 Pa and then filled with the air/hydrogen mixture. The hydrogen concentration was calculated according to the partial pressures of the gases under static experimental conditions. The films were resistively heated and their temperature was measured by a thermocouple placed close to them on the oven surface. A bias of 1V was applied, and the current through the NiO thin films was measured with a model 485 Keithley picoammeter. Current changes due to hydrogen sensing were recorded, digitized, and displayed by a computer in real time.

### 3. Results and Discussion

#### 3.1. *Structural properties*

In Figure 2 the diffractograms of four samples with different Au concentrations are presented. The Au concentration increases as the fluence of the Nd:YAG laser, which is the laser that ablates the Au target during film growth, increases. The fluence of the Nd:YAG laser for each sample is indicated in Figure 2. We observe that the NiO films crystallize mainly at the (1 1 1) and (2 0 0) orientations, which are in agreement with previous studies concerning NiO thin films grown by PLD [7]. The *average* grain size D of the NiO nanocrystallites is estimated at 46 nm by using the Scherrer equation (1):

$$D = \frac{0{,}9 \times \lambda}{FWHM \times \cos\theta} \qquad (1)$$

where $\lambda = 1{,}5418$ Angstrom ($Cu_{K\alpha}$) and $\theta$ is the central peak angle, and FWHM is the full width at half maximum of the XRD peak.

From Figure 2 it is clear that Au crystallizes as well, mainly at the (1 1 1) orientation; Au peak intensities increase and the corresponding FWHM values decrease with increasing Nd:YAG laser fluence. The reduction in the FWHM values indicates that the mean size of Au particles increases with increasing laser fluence. In particular, employing the Scherrer equation (1), the *average* size of Au nanoparticles deposited with an Nd:YAG laser fluence of 2.4 and 4.3 J/cm$^2$ is estimated at 20 and 22 nm, respectively. The size of Au particles deposited with the 1.3 J/cm$^2$ laser fluence could not be estimated using the Scherrer equation, because of the partial overlap of the two peaks corresponding to Au and NiO in the diffractogram.

### 3.2. *Morphological properties*

In Figure 3a the SEM image of a NiO film containing no Au particles ("reference sample") is shown. On this SEM image, the white spots are metallic Ni particles randomly distributed on the grey background. Figure 3b is an EDS spectrum of this background, which shows the height of the Ni peaks relative to the Si peak originating from the substrate on which the NiO film has been deposited. Figure 3c shows the EDS spectrum taken on one of the white spots shown in Figure 3a, with a remarkable increase of the Ni peaks relative to the Si peak. Such metallic droplets, consisting of the target material, are not uncommon in PLD depositions [19] and are created by insufficient oxidation of the target material inside the laser produced plasma. In Figure 3a the Ni nanoparticles are a few hundreds of nm or less in size. Nanoparticles with sizes below ca. 100 nm could not be resolved by the SEM apparatus. Therefore, there is no contradiction with the XRD results shown in Figure 2 that revealed an average NiO grain size of 46 nm. In Figure 2 we observe that there are no Ni peaks in the XRD spectra. This is due to the small amount of Ni droplets on the NiO film surface.

Figure 4a shows SEM images of the nanocomposite NiO:Au films fabricated with the lowest 1.3 J/cm$^2$ (a1) and highest 4.3 J/cm$^2$ (a2) fluence of the Nd:YAG laser, obtained with the same resolution as Figure 3a. From these images, it is clear that the number of Au nanoparticles on the film surface increases significantly with increasing laser fluence. Figure 4b shows the EDS spectrum of the grey background of the sample obtained with the highest laser fluence (a2). Comparing it with Figure 3b, it follows that Au is embedded inside the NiO matrix. From the XRD results presented above we calculated the average diameter of Au nanoparticles deposited with this laser fluence to be ca. 22 nm. Therefore, due to the restricted resolution of the SEM apparatus, they cannot be distinguished from NiO nanoparticles (46 nm average diameter). Only the larger Au nanoparticles on the film surface, with diameters of few hundreds of nm, can be identified. Figure 4c shows the EDS spectrum of such an Au nanoparticle on the film surface, where we can see the Au concentration is increased compared to the grey background shown in Figure 4b.

### 3.3. Electrical properties

The resistivity of the samples was measured using the four-point Van der Pauw method in room temperature and the results are presented in Figure 5. The resistivity of the undoped NiO film is of the same order of magnitude with what has been reported in the literature [9]. The increase in Au concentration into the NiO matrix, which follows the increase of the Nd:YAG laser fluence, causes a fast decrease of the resistivity towards conducting values. For clarity, the resistivity is fitted by an exponential dependence on the Nd:YAG laser fluence. Given the average sizes of the Au and NiO nanoparticles as they resulted from the XRD spectra above, we conclude that Au "nanoelectrodes", in view of reference [17], are embedded between the NiO nanoparticles and contribute to the strong reduction of the film resistivity.

### 3.4 $H_2$ sensing measurements

When the p-type NiO semiconductor reacts with a reducing gas like hydrogen, its resistivity increases [15]. The response $S$ of such a sensor is given as follows:

$$S = (R_g - R_o)/R_o$$

where $R_g$ and $R_o$ is the resistivity of the sensor in the gas/air mixture and in pure air, respectively. The sensor response increases with gas concentration in the gas/air mixture and in general with the sensor operating temperature. Therefore, the expected increase of the response $S$ of the NiO:Au nanocomposite films with respect to the reference NiO sample, can be practically realized in two ways: (a) for a constant gas concentration, to measure *S* for an operating temperature as low as possible, (b) for a constant operating temperature, to measure *S* for a gas concentration as low as possible. For practical reasons, we compared the sensor response of the reference NiO sample with the response of the sample with the highest Au concentration varying the operating temperature. The reference NiO sample did not respond at all to hydrogen in the 2000-10000 ppm range for operating temperatures lower than 140 $^{o}$C. Figure 6 shows the response of the NiO:Au film with the highest Au concentration to hydrogen at 117 $^{o}$C operating temperature. The response of this sensor to the presence of hydrogen is rather fast, of the order of a few minutes, and increases slightly with decreasing hydrogen concentration.

From the signal-to-noise ratio in Figure 6, it is obvious that this sensor can detect lower gas concentrations than 1000 ppm $H_2$ in air. According to our knowledge, the operating temperature and sensitivity of this NiO:Au hydrogen sensor is the optimum combination found in the literature.

In order to understand the effect of Au nanoparticles on the sensitivity of the NiO sensor, we need to consider the mechanism for hydrogen detection in air for the p-type NiO semiconductor [11]. Briefly, atmospheric oxygen is initially adsorbed on the NiO surface as $O^-$ or $O_2^-$ and thus the number of holes (majority charge carriers) increases. When $H_2$ reacts with the adsorbed oxygen on the sensor surface, the delocalized electrons recombine with the holes in the bulk and the resistivity increases, as shown in Figure 6. Based on this mechanism, the effect of Au nanoparticles on the sensor sensitivity can be described as follows: From the XRD results, we estimated the Au nanoparticles average size to be almost half the average size of the NiO crystallites. Therefore, the larger Au nanoparticles form new electrical paths between NiO nanocrystallites. This reduces the sensor resistivity dramatically (see Figure 5) and this reduction results in an increase in the sensor sensitivity [17]. Smaller Au nanoparticles are entirely lying on the NiO nanocrystallites and function as catalysts, which affect the sensor sensitivity by two different processes [20]. According to the first process, an Au nanoparticle dissociates both the $O_2$ (in the initial adsorption process) and the $H_2$ (during the sensing process) molecules and the atoms "spill over" onto the surface of the supporting semiconductor [5]. This process facilitates and speeds the reaction between the two species, according to the sensing mechanism described above. In the second process, we neglect the charged oxygen atoms adsorbed on the semiconductor and consider the adsorbed oxygen on the catalyst itself. This is the case when the growth process has resulted to a large dispersion of the catalyst nanoparticles on the semiconductor grains, a fact that results to a large increase of the effective sensor surface. The adsorbed oxygen removes electrons from the catalyst and the latter removes electrons from the supporting semiconductor. This increases the number of holes and thus reduces the sensor resistivity. Therefore, when $H_2$ reacts with the adsorbed oxygen on the catalyst, the sensor resistivity decreases again, resulting in bigger changes in the recorded signal.

**Conclusion**

Applying a versatile PLD technique with two synchronized lasers (a KrF excimer and a Nd:YAG) and two targets (Ni and Au), crystalline nanocomposite NiO:Au thin films were grown. XRD spectra show that the fluence of the Nd:YAG laser, used for the ablation of the Au target, determines the concentration and size of Au nanoparticles inside the NiO matrix. The fluence of the Nd:YAG laser also affects the effective surface morphology of the films, a fact that plays a crucial role for gas sensing. In addition, the resistivity of the films is found to strongly decrease with laser fluence. Finally, initial gas sensing measurements reveal that the NiO:Au nanocomposites can effectively detect 1000 ppm hydrogen in air and even below, at an operating temperature as low as 117 $^o$C with a fast response, resulting to an improved overall sensor performance compared to the literature.

*Acknowledgements*

*The authors would like to acknowledge the financial support of the Hellenic General Secretariat for Research and Technology through a bilateral Greek-Slovak Research Agreement (2005-07) as well as a partial support from the "Nano-structured organic-inorganic hybrid materials – synthesis, diagnostics and properties'' program No.2005ΣE01330081 of TPCI/NHRF as a "Centre of Excellence", 2005. One of the authors (I.F.) would also like to thank the TPCI/NHRF for the financial support in the frame of a 2 years scholarship.*

**Figure captions**

**Figure 1.** Schematic of the geometrical arrangement in the ablation area, showing the two lasers irradiating the Ni and Au targets.

**Figure 2.** Diffractograms of the compound thin films with increasing Au concentration (from bottom to top).

**Figure 3a.** SEM image of the reference NiO sample. **Figure 3b.** EDS spectrum of the background surface. **Figure 3c.** EDS spectrum of a Ni droplet.

**Figure 4a.** SEM images of the NiO:Au compounds with (a1) the lowest and (a2) the highest Au concentration. **Figure 4b.** EDS spectrum of the background surface in sample a2. **Figure 4c.** EDS spectrum of an Au droplet on sample a2.

**Figure 5.** Resistivity of NiO and NiO:Au thin films vs fluence of the Nd:YAG laser.

**Figure 6.** Hydrogen response of the NiO:Au thin film with the highest Au concentration, operating at 117 $^{o}$C.

Fig. 1

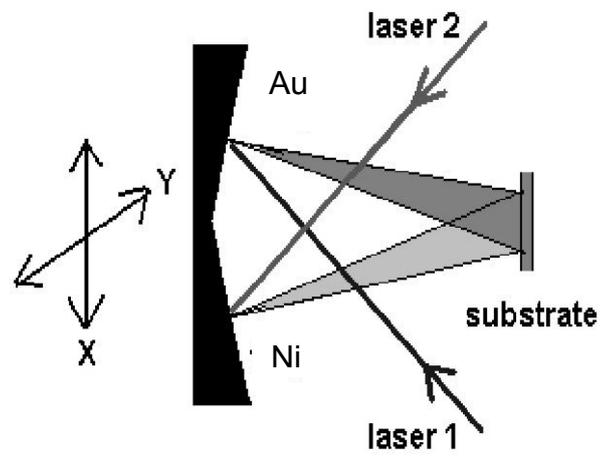

Fig. 2

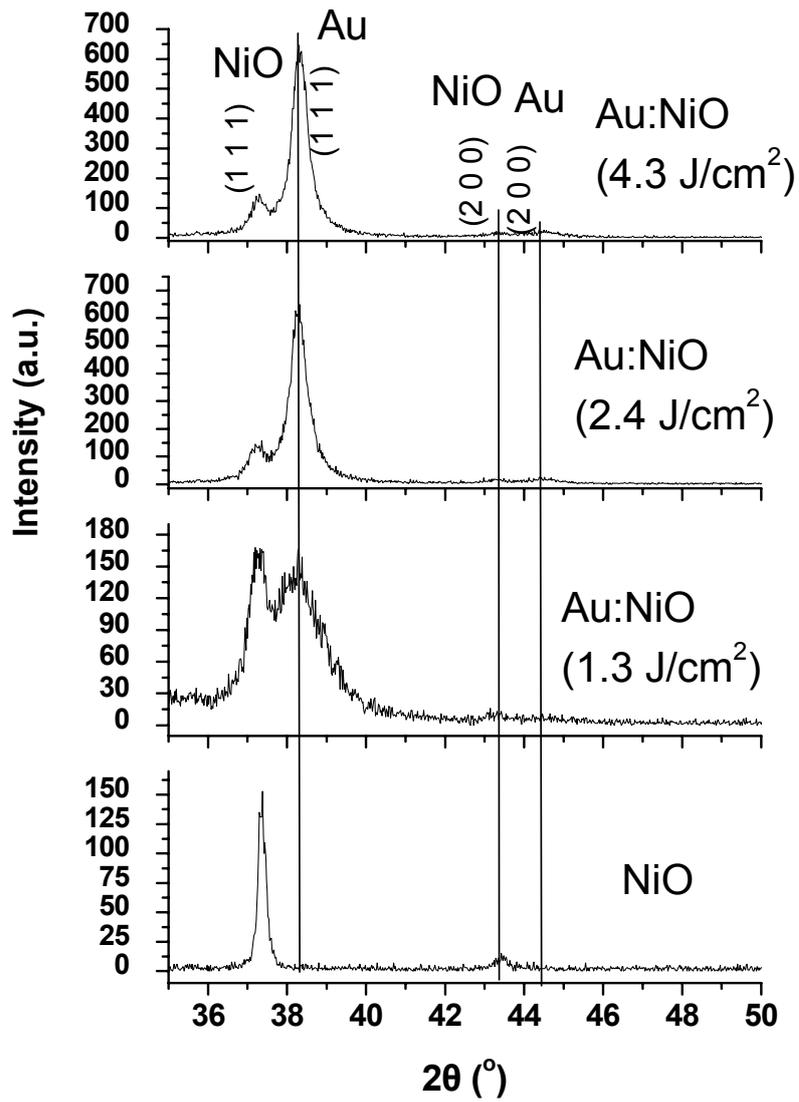

Fig. 3

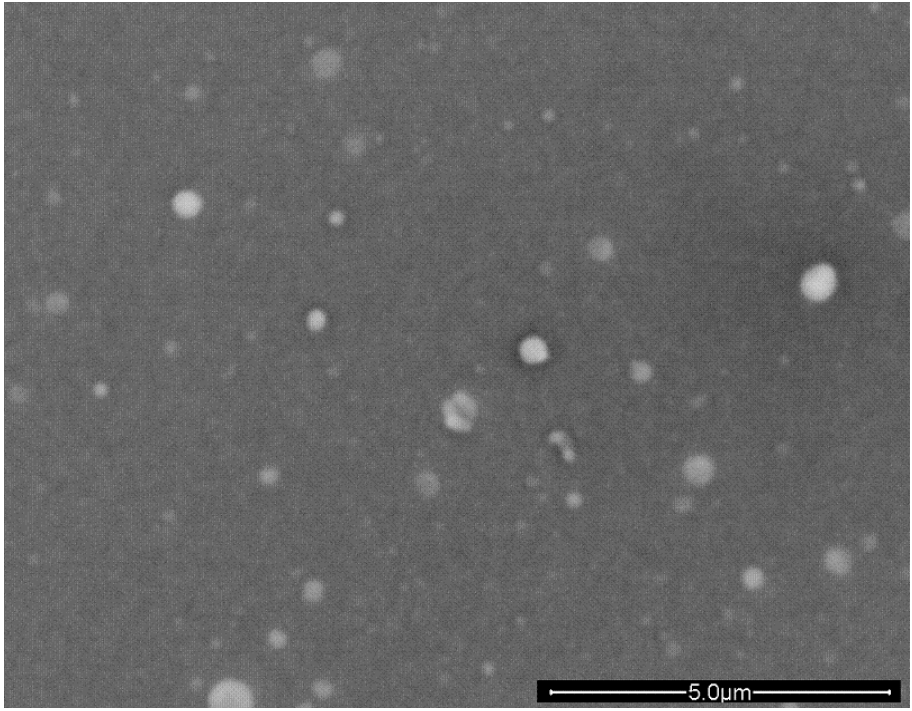

Fig. 3b and 3c

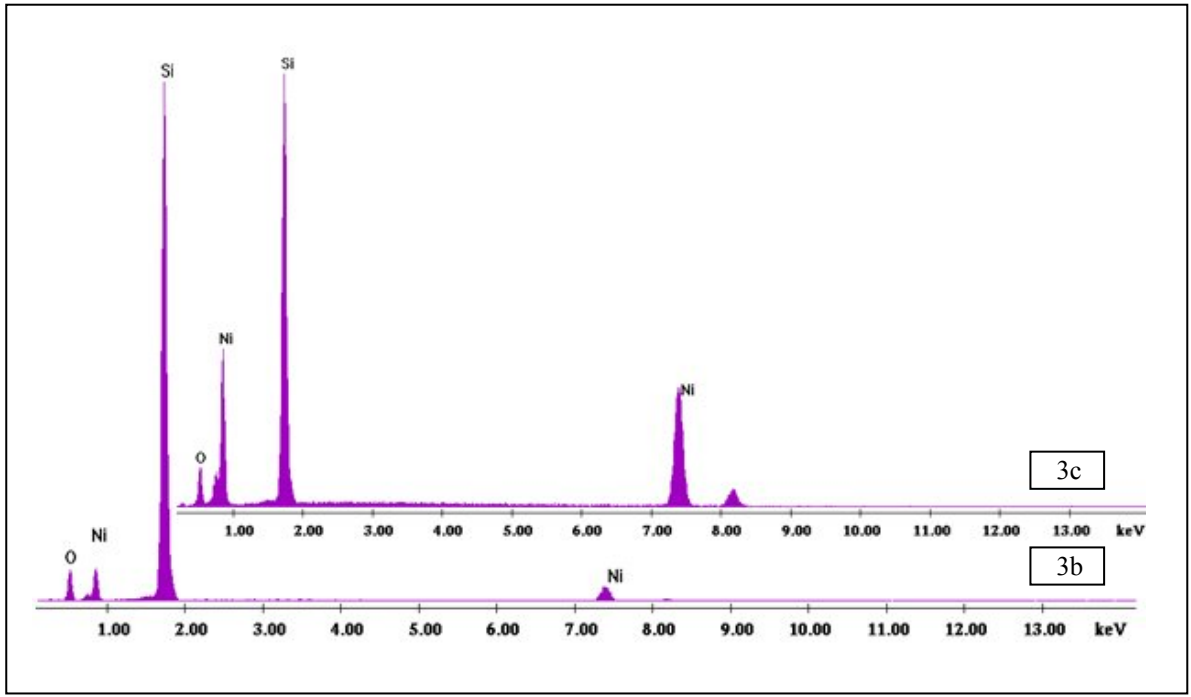

Fig. 4a

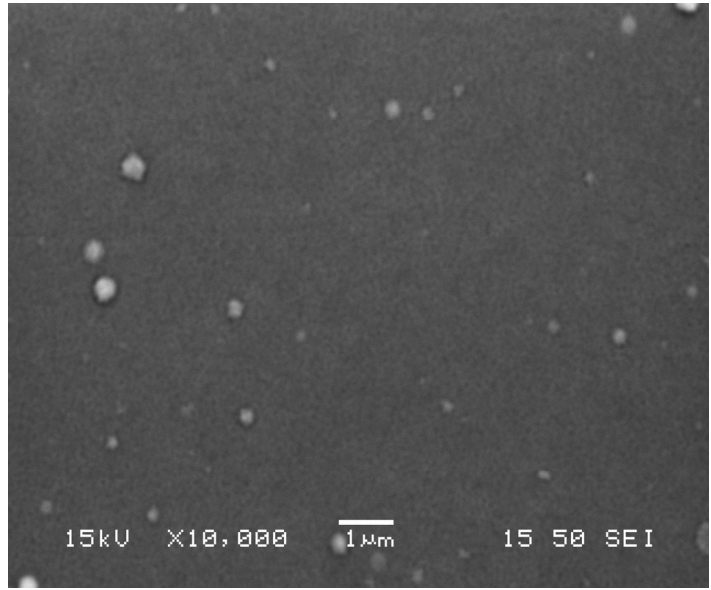

(a1)

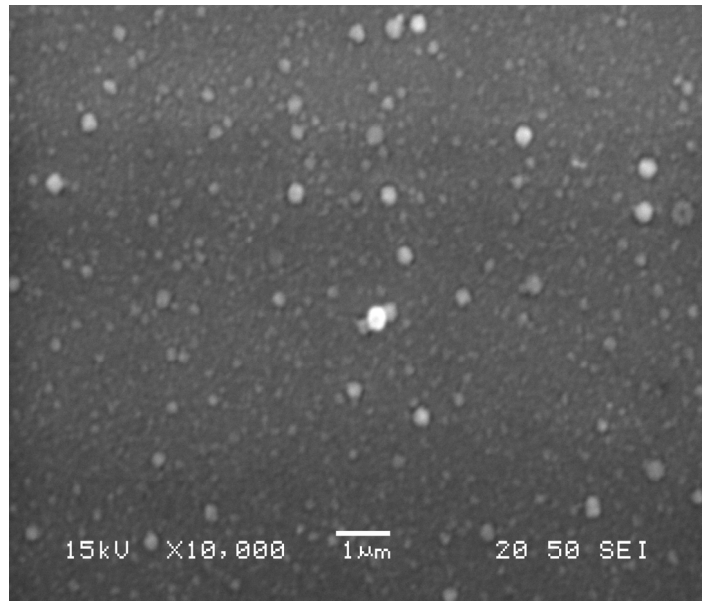

(a2)

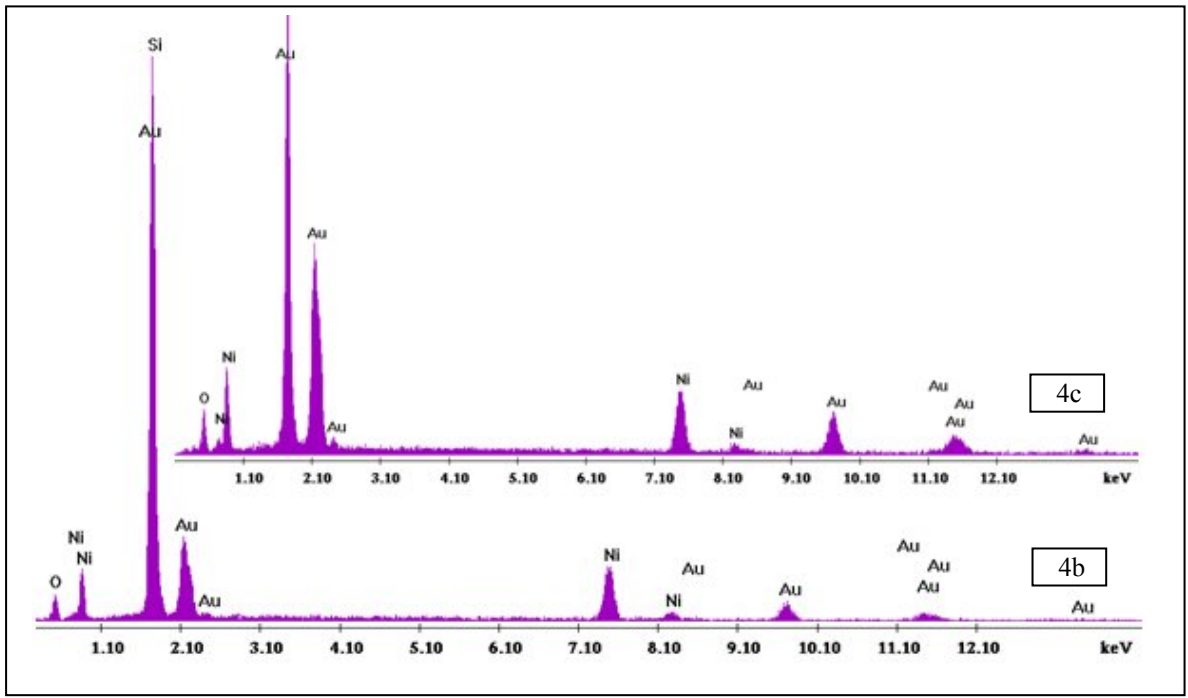

Fig. 5

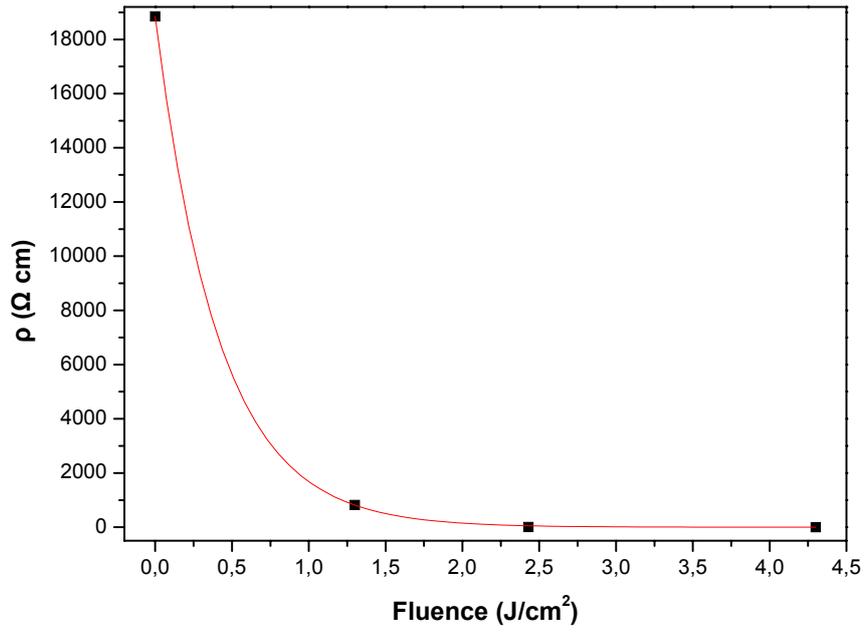

Fig. 6

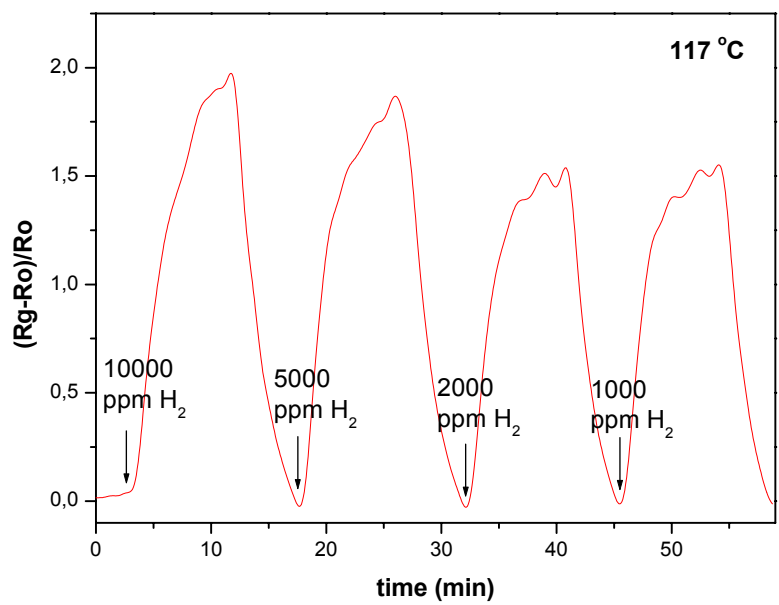